\documentclass[11pt]{article}

\usepackage[margin=1in]{geometry}
\usepackage{amsmath, amssymb, amsfonts}
\usepackage{booktabs}
\usepackage{graphicx}
\usepackage{hyperref}
\usepackage{url}

\title{Qualified Educational Capacity Planning under Heterogeneous Student
Support Needs: A Synthetic Benchmark and Decision-Support Framework}
\author{Carlos Eduardo Sanoja\\[2pt]
  Quanta Labs, LLC\\
  Professor, FCEA, Universidad Monte\'avila\\
  Edificio Lomas del Sol, Calle Humboldt, Lomas del Sol, Caracas, Venezuela\\
  \texttt{csanoja@somosquanta.com}\\
  ORCID: \href{https://orcid.org/0009-0000-0339-7072}{0009-0000-0339-7072}
  \and
  Oscar Enrique Moreno Mayz\\[2pt]
  Quanta Labs, LLC\\
  \texttt{omoreno@somosquanta.com}}
\date{\today}

\begin{document}
\maketitle

\begin{abstract}
In educational support services, the binding resource is often staff time
that is both available and \emph{qualified} for the task — and qualification
is dynamic: preparation decays, new support needs arrive that nobody is yet
prepared for, and training consumes the same staff hours that current
students need. We introduce a benchmark specification and decision-support
framework for qualified educational capacity planning. The model is a
stylized single-institution service system with heterogeneous support-demand
categories, backlog-only dynamics (services are not storable), continuous
preparation states with hard threshold qualification and decay, and
capacity-consuming training, hardened before implementation by an
adversarial multi-agent design review. The benchmark provides six
seed-controlled scenario families — announced and surprise new support
categories, staff absences, and demand surges — with exact feasibility
discipline, declared per-policy information sets, within-episode
requalification and greenfield-qualification counters, access-dispersion
metrics, replay checksums, and paired statistics. We compare service-only,
reactive, static-insurance, water-filling, and rolling-horizon mixed-integer
controllers, with an attribution chain separating service planning,
qualification maintenance, and acquisition, and a perfect-foresight
reference. Our central result is a \emph{regime map} governed by one quantity: whether a
newly required qualification can be acquired within the controller's reaction
reach (its planning horizon plus the window over which backlog stays
recoverable). When it can --- the regime that includes the core instance, and a
frozen $2{,}620$-episode adversarial suite spanning a shock-focused static plan,
disruption windows down to two periods, training rates over a $4{\times}$ range,
and low slack --- the closed-loop controller dominates ($158$ paired cells, no
static plan wins), and the attribution chain locates the value in just-in-time
qualification acquisition. When it cannot --- a cold start with deep retraining,
so the training lag exceeds the horizon --- a $4{,}420$-episode boundary search
shows lean static insurance winning by up to $2\times$ on long windows; the win
is structural, since the perfect-foresight peer ties the controller exactly and
also loses. A reactive trainer that starts after onset wastes effort and is
worst of all, and once demand structurally outruns qualified capacity no policy
choice matters. Backlog perishability, tested both post-hoc and as genuine
dynamics, shifts this boundary without erasing either regime. This contrasts
with our companion manufacturing studies, where lean static insurance holds a
regime near the capacity boundary. A transparent scenario-analysis interface, EduCapacity Studio,
reproduces any exported scenario bit-for-bit. All evidence is stylized and
synthetic; the framework makes no claims about real student outcomes,
compliance, or individual placements.
\end{abstract}

\section{Introduction}
\label{sec:p3intro}

Educational institutions plan around a resource that standard operations
models treat as fixed: qualified staff time. Yet the capacity that matters
for student support is not headcount — it is staff time that is both
\emph{available} and \emph{qualified} for the specific support task at hand,
whether that is remediation, language support, individualized learning
support, or assistive-technology assistance. Qualification is dynamic:
preparation fades without practice, new support needs arrive that nobody on
staff is yet prepared for, and the only way to create qualified capacity —
training — consumes the very staff hours that current students need. The
literature documents persistent shortages and turnover of qualified support
personnel \cite{Bettini2022QualifiedSpecialEducators,
Billingsley2019SpecialEducationAttrition}, regulatory qualification
requirements \cite{USDOEIDEAPersonnelQualifications}, and the workload
pressure on the staff who remain \cite{ASHA2002WorkloadCaseload,
Carlin2024WorkloadCaseload,Vannest2010TeacherTimeUse}; practice tools for
support-service scheduling \cite{SEATSPracticeSignal,DMSchedulesPracticeSignal,
ParaFlowToolPracticeSignal} confirm the operational need. What is missing is
a reproducible way to \emph{study} the resulting decision problem.

The decision problem is genuinely dynamic and genuinely hard. Serving
demand today and being qualified to serve demand tomorrow draw on the same
staff hours, so upskilling is never free. Qualifications decay, making
preparation a maintained asset rather than a one-off credential
\cite{Feng2013SpecialEducationTraining,Brock2015ParaprofessionalPD}.
Disruptions — a new support category, staff absences, a demand surge — can
be announced in advance or arrive as surprises, and because educational
services cannot be stored, capacity that becomes qualified too late cannot
make up for service that was never delivered. Reacting after a shock can
therefore be structurally too late, while insuring against every
contingency in advance is expensive precisely because training consumes
service time.

Existing education-operations research does not provide a testbed for this
trade-off. Timetabling benchmarks
\cite{McCollum2010ITC2,Post2014XHSTT,Ceschia2023EducationalTimetabling}
optimize courses and rooms with static staff qualifications; scheduling and
assignment decision-support systems \cite{Miranda2012UdpSkeduler,
Bailey2019StudentTeacherDSS,Aygul2026PredictPrescribeCourseScheduling} treat
qualification as input, not state; workforce-planning research outside
education models skills and training
\cite{Ernst2004StaffSchedulingReview,VanDenBergh2013PersonnelScheduling,
Czibula2017TrainerRostering} but not the education-specific combination of
support-demand categories, non-storable service, and announced-versus-
surprise information regimes; and learning-analytics dashboards
\cite{Schwendimann2017LearningDashboardReview,Kaliisa2024DashboardHype}
visualize rather than plan. We position this paper in the gap: a benchmark
specification, not a deployed scheduler.

We make four contributions. (1)~A stylized dynamic model of qualified
educational staff capacity: heterogeneous support-demand categories,
backlog-only service dynamics (services are not storable), continuous
preparation states with hard threshold qualification, decay, and training
as a capacity-consuming action — locked before implementation through a
multi-agent design review with an adversarial pass. (2)~A reproducible
synthetic benchmark: six seed-controlled scenario families spanning
announced and surprise new-category, absence, and surge shocks, with exact
feasibility discipline, declared per-policy information sets, replay
checksums, and paired statistics. (3)~A policy study comparing service-only,
reactive, static-insurance, water-filling, and rolling-horizon MILP
controllers — with an attribution chain separating service planning,
qualification maintenance, and acquisition, and an oracle reference labeled
as such. (4)~EduCapacity Studio, a transparent scenario-analysis interface
whose export/replay loop reproduces results bit-for-bit, making the
interface a reproducibility instrument rather than a recommendation engine.

The empirical message is sharper than, and instructive against, our companion
manufacturing studies. There, lean static insurance holds a regime: under
surprise shocks near the capacity boundary, reacting after onset is
structurally too late, so pre-bought cross-training wins. Here we map both sides of that
boundary and report the map as the finding. The governing quantity is whether a
newly required qualification can be acquired within the controller's reaction
reach --- its planning horizon plus the window over which backlog stays
recoverable. In the reaction-feasible regime, which includes the core instance,
the closed-loop controller dominates, and we are careful to show this is
legitimate rather than an artifact of weak comparators: the static baselines are
lean, monotone, and exactly feasible, they pre-qualify the shock category, and a
frozen adversarial suite of $2{,}620$ episodes --- a shock-focused static plan,
disruption windows down to two periods, training rates across a $4\times$ range,
and low slack --- finds no static win across $158$ paired cells. But when we push
the training lag past the controller's horizon (a cold start with deep
retraining), a $4{,}420$-episode boundary search locates the other side: lean
static insurance wins by up to $2\times$ on long windows, and the win is
structural rather than a controller defect, because the perfect-foresight peer
ties the controller exactly and also loses. Two further regimes complete the map
--- a reactive trainer that starts after onset wastes effort and is worst of
all, and under structural capacity insufficiency no policy choice matters ---
and backlog perishability shifts the boundary without erasing it. We treat this
reaction-versus-pre-positioning map, and the cross-domain contrast it draws with
our manufacturing studies, as the finding. All results are stylized
computational evidence on synthetic scenarios; the framework makes no claims
about real student outcomes, legal compliance, or individual placements.

\section{Related Work}
\label{sec:p3related}

\paragraph{Educational timetabling and scheduling benchmarks.}
Education operations research has mature, reusable benchmarks for
timetabling: the international timetabling competitions and their instance
formats \cite{McCollum2010ITC2,Post2014XHSTT,Post2016ITC3,Muller2025ITC2019},
curriculum-based formulations \cite{Bonutti2012CurriculumTimetablingBenchmark,
Fonseca2017EducationalTimetablingIP,Holm2022ITC2019MIP}, and school
timetabling surveys \cite{Pillay2014SchoolTimetablingSurvey,
Tan2021SchoolTimetablingSurvey,Ceschia2023EducationalTimetabling}. These
benchmarks optimize courses, rooms, sections, and assignments under rich
constraints, and they define the template — instances, validators, baselines
— that any new education-operations benchmark must follow. They do not model
staff qualification as a dynamic state, nor training as a decision.

\paragraph{Decision support for educational planning.}
Web-based scheduling and assignment systems are likewise established:
udpSkeduler \cite{Miranda2012UdpSkeduler}, group decision support for
academic term preparation \cite{Siddiqui2018AcademicTermGDSS},
optimization-based student-to-teacher assignment
\cite{Bailey2019StudentTeacherDSS}, and teacher--course assignment with
preferences and workload \cite{Hultberg1997TeacherAssignment,
Domenech2016TeacherAssignmentPreferences}; operational research in education
is surveyed by \cite{Johnes2015ORinEducation}. The closest dynamic threat is
predict-and-prescribe course scheduling \cite{Aygul2026PredictPrescribeCourseScheduling},
which scales course/section/location capacity from forecasts on real
university data; its capacity object is sections and rooms, not qualified
support-service staff, and upskilling never enters as a service-time
opportunity cost.

\paragraph{Workforce scheduling and training.}
Outside education, staff scheduling and multi-skilled workforce planning are
mature \cite{Ernst2004StaffSchedulingReview,VanDenBergh2013PersonnelScheduling,
Davis2018StochasticWorkforcePlanning}, including joint timetabling with
trainer rostering \cite{Czibula2017TrainerRostering} and training/service
delivery planning \cite{Senthooran2021TrainingServiceDelivery}. We adapt
these ideas — and the qualified-capacity control formulation of our
companion manufacturing studies — to an education-specific setting; the
skills-as-state, training-as-control structure is not claimed as new.

\paragraph{Qualified educational support staff.}
The practical relevance of qualification-constrained support capacity is
documented in the special-education and student-services literature:
personnel qualification requirements \cite{USDOEIDEAPersonnelQualifications},
shortages and attrition of qualified special educators
\cite{Bettini2022QualifiedSpecialEducators,Billingsley2019SpecialEducationAttrition},
certification breadth \cite{Gilmour2020CertificationArea,Kirksey2022DualCertification},
in-service training \cite{Feng2013SpecialEducationTraining,Brock2015ParaprofessionalPD,
Woulfin2021SpecialDevelopment}, paraprofessional employment
\cite{Fisher2022ParaprofessionalEmployment}, workload/caseload guidance
\cite{ASHA2002WorkloadCaseload,Carlin2024WorkloadCaseload}, and teacher time
use \cite{Vannest2010TeacherTimeUse}. Practice tools for support-service
scheduling \cite{SEATSPracticeSignal,DMSchedulesPracticeSignal,ParaFlowToolPracticeSignal}
prove the operational need while also limiting any product-novelty claim;
none provides a peer-reviewed, reproducible policy benchmark.

\paragraph{Learning analytics dashboards.}
Dashboard and advisor-analytics research is extensive
\cite{Schwendimann2017LearningDashboardReview,Matcha2020LearningAnalyticsDashboardReview,
Viberg2018LearningAnalyticsLandscape,Paulsen2024LearningAnalyticsDashboards,
Vemula2024AdvisorDashboards,Wise2019TeachingWithAnalytics} and increasingly
critical of impact claims \cite{LarrabeeSonderlund2019LearningAnalyticsInterventions,
Kaliisa2023DashboardImpact,Kaliisa2024DashboardHype,Kaliisa2023DashboardChecklist}.
This blocks interface novelty: our EduCapacity Studio is deliberately a
transparent wrapper around the benchmark — a scenario/replay layer with
declared information sets — not an analytics dashboard with outcome claims.

\paragraph{Positioning.}
Prior work covers timetabling benchmarks, educational scheduling DSSs,
assignment models, workforce planning with skills, and dashboards. What we
did not find documented is the full intersection this paper targets: a
reproducible synthetic benchmark in which heterogeneous support-demand
categories meet \emph{dynamic} qualified-staff capacity, upskilling consumes
the same staff hours as current service, demand and absence shocks come in
announced and surprise information regimes, and policy classes — including a
rolling-horizon controller — are compared under paired statistics with a
replayable decision-support interface. The contribution is that testbed and
its regime analysis, not a new timetabling solver, staffing product, or
dashboard.

\section{Problem Formulation}
\label{sec:p3formulation}

We model one educational institution (a school, campus, department, or
support center) as a dynamic service-capacity system. The scope is
deliberately stylized: planning happens at the level of support-demand
\emph{categories}, never individual students; no protected attributes, legal
compliance logic, or learning-outcome models are included; and there is no
master timetable, room, or transportation modeling. All instances are
synthetic and seed-controlled.

\paragraph{Sets and parameters.}
Periods $t \in \{0,\dots,T-1\}$ (one period is one service block; $T = 40$),
support-demand categories $g \in \mathcal{G}$, staff members
$w \in \mathcal{W}$, and qualifications $k \in \mathcal{K}$ with the identity
map $k(g) = g$ in the MVP (one required qualification per category). Static
parameters: staff hours $A_{w,t} \le H_w$ per period, qualification
thresholds $\theta_k$, training gain $\alpha_k$ per hour, decay $\delta_k$
per period, training-seat slots $cap^{\mathrm{train}}_k$, and cost
coefficients $c^B_g$ per unmet hour-period and $c^Y$ per training hour.

\paragraph{State.}
At period $t$: backlog of unmet support hours $B_{g,t} \ge 0$; staff
availability $A_{w,t}$; continuous preparation levels
$S_{w,k,t} \in [0,1]$ with hard qualification
\begin{equation}
  Q_{w,k,t} = \mathbf{1}\!\left[S_{w,k,t} \ge \theta_k\right];
  \label{eq:p3qual}
\end{equation}
and a demand-forecast window $\hat{D}_{g,t:t+F}$ whose content depends on
the scenario's information regime (announced shocks appear in the window
before onset; surprise shocks are hidden until they occur). There is no
inventory state: \emph{educational services are not storable}, so capacity
unused today cannot serve tomorrow's demand.

\paragraph{Actions and feasibility.}
Each period the planner assigns service hours
$x^{\mathrm{service}}_{w,g,t} \ge 0$ and training hours
$x^{\mathrm{train}}_{w,k,t} \ge 0$ subject to
\begin{align}
  \sum_g x^{\mathrm{service}}_{w,g,t} + \sum_k x^{\mathrm{train}}_{w,k,t}
    &\;\le\; A_{w,t} \quad \forall w,
  \label{eq:p3budget}\\
  x^{\mathrm{service}}_{w,g,t} \;\le\; A_{w,t}\, Q_{w,k(g),t},
  \qquad
  \bigl|\{w : x^{\mathrm{train}}_{w,k,t} > 0\}\bigr|
    &\;\le\; cap^{\mathrm{train}}_k \quad \forall k.
  \label{eq:p3elig}
\end{align}
Constraint~\eqref{eq:p3budget} is the central mechanism: upskilling consumes
the same scarce staff hours that direct service needs now. Eligibility is
hard --- only qualified staff can serve a category.

\paragraph{Dynamics.}
Realized demand is the scenario mean with multiplicative truncated noise,
gated to active categories. Served hours and the backlog-only queue evolve
as
\begin{align}
  y_{g,t} = \sum_w x^{\mathrm{service}}_{w,g,t},
  \qquad
  B_{g,t+1} &= \max\!\bigl(0,\; B_{g,t} + D_{g,t} - y_{g,t}\bigr),
  \label{eq:p3backlog}\\
  S_{w,k,t+1} = \Pi_{[0,1]}\!\bigl((1-\delta_k)\, S_{w,k,t}
    + \alpha_k\, x^{\mathrm{train}}_{w,k,t}\bigr),
  \qquad
  Q_{w,k,t+1} &= \mathbf{1}\!\left[S_{w,k,t+1} \ge \theta_k\right].
  \label{eq:p3skill}
\end{align}
Decay makes qualification a maintained asset: an unrefreshed qualification
eventually lapses and can only be recovered by further training.

\paragraph{Objective and information regimes.}
The per-period cost is
$c_t = c^B \sum_g B_{g,t+1} + c^Y \sum_{w,k} x^{\mathrm{train}}_{w,k,t}$,
and a policy is a causal map from the observation to actions minimizing
$\sum_t c_t$ under the scenario's disruption process (new support
categories, staff absences, and demand surges, each in announced and
surprise variants). Every policy declares its information set explicitly,
and oracle references are labeled as such. The benchmark's decision tension
is intertemporal: serving demand today and being \emph{qualified} to serve
demand tomorrow draw on the same staff hours
through~\eqref{eq:p3budget}, while hard
qualification~\eqref{eq:p3qual} makes future capacity a discrete,
training-lagged consequence of today's allocation --- and because services
cannot be stored~\eqref{eq:p3backlog}, capacity that arrives too late
cannot be made up in advance.

\section{Benchmark Design}
\label{sec:p3benchmark}

The benchmark is a reproducible, seed-controlled environment realizing the
formulation of Section~\ref{sec:p3formulation}, hardened before
implementation by a multi-agent design review (two specialist proposals
reconciled by an adversarial pass that rejected, among others, a decay rate
that degenerated the no-shock control and a category count that was
infeasible by construction).

\paragraph{Frozen instance.}
Eight staff members with $8$ hours per period; $G \in \{3, 6, 9\}$ support
categories with $G = 6$ as the default and the category-count sweep holding
the slack ratio constant ($\bar{D}_g = 64/(1.5\,G)$) so that coverage
differences come from qualification breadth rather than raw overload;
$\theta = 0.6$, $\alpha = 0.08$ (one cold-start qualification costs
$7.5$ training hours, roughly one staff-period), $\delta = 0.01$;
per-category demand mean $8.0$ hours (slack ratio $1.33$) with
multiplicative truncated noise ($\sigma = 0.2$); three training seats per
qualification per period. The deterministic initial qualification matrix is
lean on purpose: two home categories per staff member at $S_0 = 0.8$,
everything else warm but unqualified ($0.25$), and the latent shock
category unqualified for everyone ($0.2$) --- so a new support category
can never be served without prior training.

\paragraph{Scenario families and information regimes.}
Six seed-controlled families: a stationary no-shock control; a new support
category activating mid-episode (demand $12$ h for $20$ periods, onset
randomized per seed) in \emph{announced} (visible once the $F{=}4$ forecast
window reaches the onset) and \emph{surprise} (hidden until onset) variants;
staff absence (three of eight staff unavailable for eight periods) in
announced and surprise variants; and a demand surge ($\times 1.75$ for ten
periods) that pushes the system transiently over capacity. Realized onsets
are logged for replay.

\paragraph{Metrics.}
Operational: total unmet support hours, mean coverage ratio
($\min(1, y/(D{+}B))$ over active category-periods — the storable-goods
notion of fill rate does not apply), peak and area-under backlog, staff
utilization. Resilience: recovery rate and time (backlog within half a
period of demand of its pre-shock level for two consecutive periods),
unrecovered counts. Capability: within-episode requalifications and
greenfield qualifications (first-ever threshold crossings), training hours
by staff and qualification. Access dispersion: Jain indices over
per-category coverage and per-staff training hours plus minimum floors ---
reported strictly as coverage/training dispersion; no demographic
attributes are modeled and no fairness claims are made. Reproducibility:
seeds, configs, solver status and runtime, fallback counts, and a replay
checksum over the backlog/qualification trajectory.

\paragraph{Feasibility discipline.}
The environment validates every action and repairs infeasibility
deterministically while counting violations and qualification-zeroed hours;
all policies in this paper emit exactly feasible actions, so these
diagnostics are identically zero in every reported run.

\paragraph{Decision-support interface.}
EduCapacity Studio, a research demonstrator, exposes the benchmark through
seven screens (scenario builder; category, staffing, and training editors;
policy comparison with each policy's declared information set; a bottleneck
explorer; and scenario replay/export). Exported scenarios store the
configuration, seed, and replay checksums; re-importing re-runs the
simulator and verifies identical results, making the interface itself a
reproducibility check rather than a recommendation engine. It is not a
scheduling, compliance, or student-placement tool.

\section{Policies and the Rolling-Horizon Controller}
\label{sec:p3controller}

\paragraph{Baseline policy classes (explicit information sets).}
\textbf{ServiceOnly} sees the current backlog, nowcast demand,
qualifications, and availability; it never trains and allocates service
hours proportionally to outstanding demand (a water-filling split — the
serve-scarcest-to-exhaustion rule was rejected because it oscillates near
tight capacity). \textbf{ReactiveGap} additionally trains toward a
category's qualification only after its uncovered demand exceeds a
threshold. \textbf{StaticInsurance$\{b\}$} executes a fixed lean
cross-qualification plan from $t=0$ using only the $t=0$ configuration,
spending a budget fraction $b \in \{0.05, 0.10, 0.20\}$ of total staff hours
spread over half the horizon ($b = 0.20$ is a deliberately saturating
endpoint whose per-period diversion exceeds typical slack, documented as
such). \textbf{WaterFillingTraining} spreads a $10\%$ training budget across
observed qualification gaps weighted by visible demand.
\textbf{OracleUpperReference} runs the \emph{same} receding-horizon controller
($H{=}6$) as the primary but with surprise activations unmasked in its forecast
— perfect demand foresight, privileged information, clearly labeled. It is a
perfect-foresight \emph{peer} reference, not a global upper bound (the name is
historical): on announced shocks it is informationally identical to the primary
and may tie or be edged by it within solver tolerance, and it never competes in
the policy tables.

\paragraph{ForecastAwareMPC (primary).}
At every period the controller observes the state, builds demand and
availability forecasts from the observation window (surprise shocks are
masked by the environment until onset; announced absences expose the true
window, surprise absences only persistence), solves a finite-horizon
mixed-integer program, applies only the first-period action, and replans.
The prediction model uses hard observed qualification at $h=0$ (so the
executed action is exactly feasible), binary predicted qualification
$\theta\, c_{w,k,h} \le S_{w,k,h}$ for $h \ge 1$, the shared time
budget~\eqref{eq:p3budget}, seat caps (hours relaxation inside the program,
slot trim on the executed action), and the backlog queue encoded as
$B_{h+1} \ge B_h + \hat{D}_h - y_h$, $B \ge 0$, which is tight at the
optimum because backlog costs are strictly positive; realized backlog is
recomputed by the environment's true $\max(0,\cdot)$ recursion. The
terminal value prices qualification gaps left open at the horizon edge,
\begin{equation}
  gap_k = \max\!\bigl(0,\; \widehat{sd}_k - \textstyle\sum_w
    \hat{A}_{w,H-1}\, c^T_{w,k}\bigr),
  \qquad
  V_f = \lambda_{\mathrm{gap}} \sum_k gap_k ,
  \label{eq:p3gap}
\end{equation}
with $\widehat{sd}_k$ the per-qualification maximum of visible forecast
demand from the horizon edge onward and binary terminal qualifications
$\theta\, c^T_{w,k} \le S_{w,k,H}$. The \emph{primary configuration},
locked ex ante before any validation run, is $H = 6$,
$\lambda_{\mathrm{gap}} = 1.5$; the $\lambda \in \{0, 1.5, 5\}$ and horizon
sweeps are sensitivity analyses, never best-of selection. Attribution
ablations mirror the chain used in our companion controller study:
\textbf{ServiceOnlyMPC} (receding-horizon service allocation without
training variables) and \textbf{MaintenanceMPC} (training restricted to
currently held qualifications, which therefore can maintain but never
acquire or recover a qualification), so the full controller's value
decomposes into service planning, qualification maintenance, and
acquisition.

\paragraph{Solver and diagnostics.}
Each replanning step solves with \texttt{scipy.optimize.milp} (HiGHS branch
and bound) under a one-second limit and a $10^{-2}$ relative gap — replanning
corrects residual suboptimality — with binaries reduced by fixing
$c = 1$ for qualification cells that cannot decay below threshold within the
horizon. Every solve logs status and wall-clock time; a failed solve would
fall back to WaterFillingTraining and be counted, and fallbacks are zero in
all accepted runs.

\section{Experimental Setup}
\label{sec:p3setup}

\paragraph{Protocol.}
All experiments are synthetic and seed-controlled on the frozen instance of
Section~\ref{sec:p3benchmark}; no real institutional or student data are
used anywhere. The primary controller configuration (ForecastAwareMPC,
$H{=}6$, $\lambda_{\mathrm{gap}}{=}1.5$) was locked ex ante; horizon and
$\lambda$ variations are sensitivity analyses. Twenty paired seeds per
validation cell (the oracle reference uses eight, each a perfect-foresight
receding-horizon $H{=}6$ solve, not a full-horizon program); a smoke suite at
$T{=}20$ gates the implementation with
eight checks including exact feasibility, deterministic replay across all
scenario--policy pairs, announced-vs-surprise distinguishability over
multiple seeds, structural failure of ServiceOnly on the new-category
scenarios, and a no-pre-onset-leakage regression for surprise shocks.

\paragraph{Experiment families.}
(i) The six core scenarios against the seven-policy set; (ii) the
attribution chain (ServiceOnlyMPC, MaintenanceMPC) on no-shock, announced
and surprise new-category, and announced absence; (iii) the
$\lambda$ ablation on the new-category scenarios; (iv) a slack sweep
(demand $10.16/8.0/5.33$ hours per category, i.e.\ slack ratios
$1.05/1.33/2.0$) on the surprise new category; (v) a training-speed sweep
($\alpha \in \{0.04, 0.08, 0.16\}$) on the announced new category,
including short-horizon $\lambda$ cells in the slow regime where the
training lag approaches the horizon; (vi) the category-count sweep
($G \in \{3, 6, 9\}$ at constant slack) probing where lean static insurance
stops scaling; (vii) the oracle reference on three shocked scenarios;
(viii)~a frozen \emph{adversarial threat-closure} suite ($2{,}620$ episodes)
that adds a shock-focused static comparator and grids training speed
$\alpha \in \{0.02, 0.04, 0.08, 0.16\}$ against disruption duration
$\in \{2, 4, 8, 12, 20\}$ under surprise, plus low-slack/slow-training cells;
(ix)~a post-hoc \emph{deadline/perishability} re-scoring of the executed
episodes at $L \in \{1, 2, 4, \infty\}$; and (x)~a frozen \emph{boundary
search} ($4{,}420$ episodes) that locates where pre-positioning beats reaction,
gridding training speed $\alpha \in \{0.01, 0.02, 0.04, 0.08\}$ against
disruption duration with cold and warm shock starts, a static backup-count
sweep, the perfect-foresight Oracle peer in every cell, a genuine
perishable-dynamics variant (\texttt{perishable\_env}, deadline $L$, lost
penalty $c_{\mathrm{lost}}$), and a structural-insufficiency sweep
($\rho \in \{1.33, 0.89, 0.67, 0.53\}$). Families (viii)--(x) are reported
separately under \texttt{experiments/} (\texttt{adversarial\_validation\_*},
\texttt{deadline\_backlog\_*}, \texttt{boundary\_search\_*}) and are never
pooled with the core tables.

\paragraph{Statistics and reproducibility.}
Policy comparisons are paired by seed: per-seed win rates with exact
two-sided sign tests, and separately a seeded paired bootstrap
(10{,}000 resamples) confidence interval on the mean cost difference with
relative effect sizes. The two criteria are not interchangeable; borderline
results are described as mean-effect evidence rather than decisive win-rate
evidence. Episodes replay deterministically: each run records a checksum
over the backlog/qualification trajectory, and two independent executions
of the full suite produce byte-identical artifacts once wall-clock solver
columns are excluded. Every policy row reports solver status, mean solve
time, fallback counts (zero throughout), and the environment's
repair/eligibility diagnostics (zero throughout).

\section{Results}
\label{sec:p3results}

We report results by mechanism and regime, and the headline is a \emph{regime
map} rather than a ``method wins'' claim. In the reaction-feasible regime that
includes the core instance --- where a newly required qualification can be
acquired within the controller's planning horizon and recoverable window --- the
closed-loop controller dominates, and we establish this is a legitimate
consequence of problem structure (the controller coincides with its
perfect-foresight peer and the static baselines are lean and functional) rather
than weak comparators. We then \emph{locate the boundary}: when the training lag
exceeds the controller's reach, lean static insurance wins; a reactive trainer
that starts too late wastes effort; and under structural capacity insufficiency
no policy choice matters.

\subsection{Implementation and reproducibility}
\label{sec:p3res-impl}

The validation suite runs 1{,}544 episodes across 79 cells. Every policy is
exactly feasible: zero environment repairs and zero qualification-zeroed
service hours across all rows (including the oracle reference). The MILP
controllers run a node-limited branch-and-bound made deterministic by
single-threaded execution (set via \texttt{OMP\_NUM\_THREADS=1}; the HiGHS
\texttt{threads} option is also forwarded, though the SciPy wrapper reports it
as unrecognized); determinism is \emph{verified}, not merely asserted, by a
replay-checksum spot-check over the hardest MPC cells, which passes here and in
both the adversarial and deadline suites. No NaNs or negative states occur.

\subsection{A perfect-foresight reference and the value of foresight}
\label{sec:p3res-oracle}

Table~\ref{tab:p3oracle} compares the primary controller against a
perfect-foresight reference: the same receding-horizon controller with
surprise activations unmasked in its forecast. We use it to \emph{isolate the
value of foresight}, not as a global optimum --- it is a peer controller, not
an upper bound, and on announced shocks it is informationally identical to the
primary (which already sees the announced shock), so the two coincide to
within solver tie-breaking tolerance ($\pm1$--$2\%$, including a case where
the primary edges it). The informative cell is the surprise shock, where the
primary is genuinely information-limited: foresight is worth a $24\%$ cost
reduction there. The legitimacy of the dominance results below therefore does
\emph{not} rest on this reference; it rests on the baseline-fairness and
adversarial-robustness evidence of Sections~\ref{sec:p3res-core}
and~\ref{sec:p3res-regime}.

\begin{table}[t]
\centering
\caption{Primary vs.\ the perfect-foresight reference (mean cost; 8 seeds,
the reference's budget). On announced shocks the two coincide within solver
tolerance (foresight adds nothing once the shock is announced); the surprise
gap isolates the value of foresight.}
\label{tab:p3oracle}
\small
\begin{tabular}{lrrr}
\toprule
Scenario & Primary & Perfect-foresight & gap \\
\midrule
announced new category & 335 & 339 & $-1\%$ \\
surprise new category  & 419 & 339 & $+24\%$ \\
announced absence      & 670 & 657 & $+2\%$ \\
\bottomrule
\end{tabular}
\end{table}

\subsection{Core regimes}
\label{sec:p3res-core}

Table~\ref{tab:p3core} reports mean cost across the six core scenarios for
the lean, functional baselines and the primary controller. Three patterns
hold. First, the service-only policy collapses wherever qualification binds:
on the new-category scenarios no staff is initially qualified, so its shock
coverage is zero and cost is an order of magnitude above the controller.
Second, every training-capable policy improves on service-only, and the
static-insurance plans are genuinely lean here --- they pre-qualify backups
for the latent shock category (shock coverage $0.55$--$0.56$, $6$--$11$
greenfield qualifications, $53$--$95$ training hours) rather than the
self-defeating blanket over-training of a misconfigured plan. Third, the
closed-loop controller nonetheless wins every regime, including surprise
shocks, low slack, slow training, and the category-count sweep.

\begin{table}[t]
\centering
\caption{Core scenarios (mean cost, 20 seeds). ST10 = StaticInsurance (10\%
backup target), WF = WaterFillingTraining. All comparisons Primary vs.\ each
baseline are 20--0 on paired seeds (sign $p = 1.9\times10^{-6}$), with
bootstrap CIs excluding zero.}
\label{tab:p3core}
\small
\begin{tabular}{lrrrrr}
\toprule
Scenario & ServiceOnly & ReactiveGap & ST10 & WF & Primary \\
\midrule
no\_shock              & 3{,}333 & 345   & 2{,}450 & 650   & \textbf{164} \\
announced new category & 9{,}319 & 1{,}635 & 3{,}602 & 1{,}741 & \textbf{335} \\
surprise new category  & 9{,}319 & 1{,}635 & 3{,}602 & 1{,}845 & \textbf{419} \\
announced absence      & 5{,}510 & 2{,}209 & 3{,}628 & 2{,}137 & \textbf{670} \\
surprise absence       & 5{,}510 & 2{,}209 & 3{,}628 & 2{,}137 & \textbf{699} \\
demand surge           & 8{,}600 & 7{,}240 & 6{,}599 & 5{,}678 & \textbf{3{,}035} \\
\bottomrule
\end{tabular}
\end{table}

\subsection{Attribution: where the value comes from}
\label{sec:p3res-attrib}

The ablation chain (Table~\ref{tab:p3attrib}) decomposes the controller's
advantage on the announced new-category shock, holding the receding-horizon
service allocation constant. Replacing the heuristic service split with the
MILP allocation (ServiceOnly $\to$ ServiceOnlyMPC) accounts for a $27\%$
reduction; adding qualification \emph{maintenance} against decay
(ServiceOnlyMPC $\to$ MaintenanceMPC) roughly halves cost again; and enabling
\emph{acquisition} of the new qualification just in time
(MaintenanceMPC $\to$ Primary) removes $91\%$ of the remaining gap. The
dominant mechanism is therefore targeted just-in-time qualification, not
merely better allocation --- which is precisely the capability static
insurance approximates but pays for in advance.

\begin{table}[t]
\centering
\caption{Attribution chain on the announced new-category shock (mean cost,
20 seeds), holding service allocation constant across the MPC variants.}
\label{tab:p3attrib}
\small
\begin{tabular}{lrl}
\toprule
Policy & cost & mechanism added \\
\midrule
ServiceOnly      & 9{,}319 & --- (heuristic service, no training) \\
ServiceOnlyMPC   & 6{,}793 & MILP service allocation \\
MaintenanceMPC   & 3{,}770 & + qualification maintenance \\
Primary          & 335   & + just-in-time acquisition \\
\bottomrule
\end{tabular}
\end{table}

\subsection{A regime map: where reaction beats pre-positioning, and where it does not}
\label{sec:p3res-regime}

The companion manufacturing studies find a regime split: lean static insurance
wins under surprise shocks near the capacity boundary, where a reaction
transient is structurally unrecoverable. The scientific question is not whether
the controller \emph{always} wins here, but \emph{where} the boundary lies. The
governing quantity is whether a newly required qualification can be acquired
within the controller's \emph{reaction reach} --- its planning horizon $H{=}6$
plus the window over which backlog stays recoverable. We map both sides of that
boundary and treat the map, not a dominance claim, as the finding.

\paragraph{Reaction-feasible regime (the controller wins).}
In the core instance the disruption window is long ($20$ periods) relative to a
short training lag ($1$--$2$ periods), and non-storable backlog is recoverable
over such a window, so the controller reacts within the disruption and the
surprise/announced distinction that static insurance hedges largely dissolves.
The sweeps confirm the gap narrows but does not close: at low slack the relative
margin shrinks (Primary $4{,}115$ vs.\ ST10 $9{,}700$, a $2.4\times$ gap versus
$8\times$ at high slack), and at the slow training rate the controller still
wins $20$--$0$. To stress this side of the boundary we
ran a dedicated, frozen adversarial threat-closure suite
(\texttt{adversarial\_validation\_*}; $2{,}620$ episodes, zero
fallbacks, zero repairs, deterministic spot-check passing) attacking the result
along the four axes a reviewer would press --- short disruption windows, slow
training, low slack, and a stronger static comparator. The comparator set adds
\textbf{ShockFocusedStatic}, a legitimate exact-feasible plan that concentrates
its entire slack-metered, seat-capped backup budget on the latent shock
qualification from $t=0$ (it pre-commits to the shock \emph{slot} but never
learns its timing) --- one of several static variants spanned by a backup-count
sweep. Across all $158$ paired
Primary-vs-comparator cells, no static plan beats the controller. Three
representative cells: (i)~the shock-focused plan under surprise $+$ low slack
$+$ slow training ($\alpha{=}0.04$) loses $20$--$0$ (Primary $4{,}159$ vs.\
$14{,}254$ --- concentrating the premium on the guessed slot starves current
service); (ii)~a two-period window with the training lag deliberately made
\emph{longer} than the window ($\alpha{=}0.02$, lag $\approx 3$ periods), the
canonical case where pre-positioning should win, loses $8$--$0$ (Primary $263$
vs.\ StaticInsurance10 $2{,}737$ --- the controller absorbs and serves down a
short backlog rather than pre-training for a shock that barely materializes);
and (iii)~a grid over training speed $\alpha \in \{0.02, 0.04, 0.08,
0.16\}$ and disruption duration $\in \{2, 4, 8, 12, 20\}$ in which the
controller wins all $20$ cells. These cells share warm starts and reskilling
rates at which the training lag fits inside the controller's reach, so they all
sit in the reaction-feasible regime.

\begin{figure}[t]
\centering
\includegraphics[width=\linewidth]{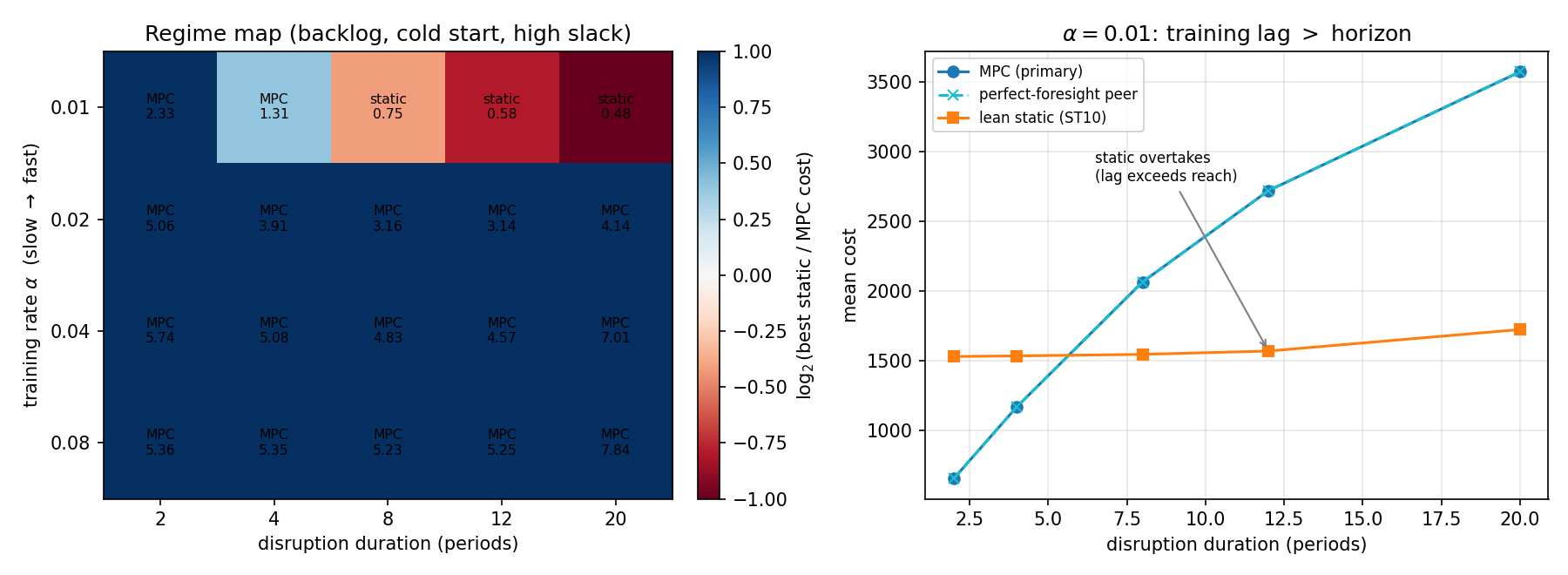}
\caption{The reaction-versus-pre-positioning boundary (\texttt{boundary\_search\_*}).
Left: winner over training rate $\alpha$ (slow at top) and disruption duration,
backlog-only with a cold start; the static-favoring corner (slow training, long
window) is where the qualification lag outruns the controller's reach. Right: at
$\alpha{=}0.01$ the controller and its perfect-foresight peer coincide
\emph{exactly} (the lag exceeds the $H{=}6$ horizon, so foresight cannot help),
while lean static insurance, flat in duration, overtakes both once the window is
long enough to amortize its premium.}
\label{fig:p3regime}
\end{figure}

\paragraph{Pre-positioning regime (static insurance wins).}
To find the other side of the boundary we ran a second frozen suite
(\texttt{boundary\_search\_*}; $4{,}420$ episodes, zero fallbacks,
zero repairs, determinism spot-check passing) that pushes the training lag past
the controller's horizon: a \emph{cold} shock start (no prior preparation) and
\emph{deep} retraining ($\alpha{=}0.01$, so a new qualification costs
$\approx\!60$ staff-hours --- a lag of $\approx\!7$--$8$ periods), with ample
slack so pre-positioning is affordable. Here a \emph{lean} static plan
(StaticInsurance$10$) wins once the window is long enough to accumulate costly
backlog: it beats the controller $0$--$20$ (sign $p{=}1.9\times10^{-6}$) at every window
$\ge\!8$ periods, by up to $2\times$ at duration $20$ (Primary $3{,}570$ vs.\
$1{,}725$). The win is \emph{structural}, not a baseline or information artifact:
the perfect-foresight Oracle --- the same $H{=}6$ controller with the shock
unmasked --- ties the primary \emph{exactly} ($0$--$0$--$20$, mean difference
$0.0$) and loses to static identically, because no $H{=}6$ controller can
accumulate $60$ training-hours within a six-period horizon. At $\alpha\ge0.02$
(lag inside the horizon) the ordering flips back: foresight pays (the Oracle
beats the primary) and the primary beats static. The boundary is therefore set
by the training lag relative to the controller's reach (Figure~\ref{fig:p3regime});
a static commitment from
$t=0$ --- which needs the shock \emph{slot} but never its timing --- is the only
policy that pre-qualifies in time when that lag is long. The heavier static
plans over-train and lose: lean pre-positioning is what wins.

\begin{table}[t]
\centering
\caption{Regime map. Which wins --- just-in-time reaction (the controller) or
advance pre-positioning (static insurance) --- is governed by the
new-qualification training lag relative to the controller's reaction reach
(horizon $H{=}6$ plus the recoverable/perishable window). All four regimes are
reproduced in frozen suites.}
\label{tab:p3regime}
\small
\begin{tabular}{@{}l p{0.46\linewidth} l@{}}
\toprule
Regime & Condition & Best policy \\
\midrule
Reaction-feasible & lag $\le$ reach (warm start; reskilling $\alpha\ge0.02$; or short window) & controller \\
Pre-positioning & lag $>$ reach (cold start, deep retrain $\alpha{=}0.01$) and long window ($\ge8$) & lean static insurance \\
Reaction-too-late & reactive trainer starts after onset; cannot qualify in time & none (wasted training) \\
Structurally insufficient & demand $>$ qualified capacity ($\rho<1$) & none (all collapse) \\
\bottomrule
\end{tabular}
\end{table}

\paragraph{Reaction-too-late and structurally-insufficient regimes.}
Two further regimes complete the map. A purely reactive trainer (ReactiveGap),
which trains only after the gap appears, is the \emph{worst} policy in the
pre-positioning regime: at $\alpha{=}0.01$, duration $20$ it burns $1{,}100$
training-hours after onset, too late to qualify (shock coverage $0.01$), and
underperforms even no-training ($8{,}404$ vs.\ ServiceOnly $6{,}877$) ---
training that starts too late is worse than not training. And when demand
structurally outruns total qualified capacity ($\rho<1$), every policy collapses
to uniformly poor coverage and the choice stops mattering: as $\rho$ falls from
$1.33$ to $0.53$, mean coverage drops from $0.88$ to $0.12$ and the worst
comparator's relative gap over the controller shrinks from $+2{,}118\%$ to
$+20\%$. The terminal qualification-gap penalty has a small
effect at this fast training rate (Primary vs.\ $\lambda{=}0$: $15$--$5$ on
the announced shock, $13$--$7$ on surprise, mean effect $14\%$ on announced-shock cost),
consistent with just-in-time training being near-optimal when the lag is
short; we report it as a sensitivity rather than a decisive lever and do not
pool it across rates.

\subsection{Backlog perishability moves the boundary}
\label{sec:p3res-deadline}

The backlog-only dynamics treat uncovered support as recoverable later, which a
reviewer can fairly attack: perhaps the controller ``wins'' only by serving late
what an institution cannot ethically or operationally defer. We test this with a
frozen deadline re-scoring (\texttt{deadline\_backlog\_*}): each executed
trajectory is re-evaluated under a deadline $L$, counting support
served more than $L$ periods after it arrived as late/lost rather than
recovered. This is an explicit \emph{post-hoc} lens, not a re-optimization ---
policies are not re-planned against the deadline --- so it isolates whether the
advantage \emph{depends on} late service; $L{=}\infty$ reproduces the
backlog-only result exactly. It does not depend on late service: the controller
loses the fewest hours at every deadline $L \in \{1, 2, 4, \infty\}$ across all
four tested scenarios, $20$--$0$ paired in every cell. Even at the strictest
deadline $L{=}1$ on the announced new-category shock, the controller leaves $20$
hours undelivered against $540$--$680$ for the static plans (on-time coverage
$0.99$ vs.\ $0.69$--$0.75$). In the reaction-feasible regime the controller's
lead therefore does not rest on forgiving backlog recovery. But perishability is
not neutral at the boundary. We also re-ran the boundary suite under
\emph{genuine} perishable dynamics (\texttt{perishable\_env}: support unserved
past a deadline $L$ is permanently lost at penalty $c_{\mathrm{lost}}{=}20$,
$L\in\{2,4\}$). A hard deadline is double-edged: it penalizes the controller's
deferred shock service --- which, in the pre-positioning regime, it never
delivers --- but also penalizes the static plan's starvation of base categories
while it pre-trains. The net effect moves the boundary without erasing either
regime: at the cold, deep-retraining corner the static-win region \emph{persists
but contracts to the longest windows} --- static wins at duration $20$ (Primary
$5{,}330$ vs.\ $4{,}436$ at $L{=}2$; $5{,}721$ vs.\ $3{,}794$ at $L{=}4$, both
$0$--$20$) while the controller reclaims the shorter eight-period windows. A
deadline-aware controller is the natural response in that regime and is left to
future work.

\begin{figure}[t]
\centering
\includegraphics[width=0.92\linewidth]{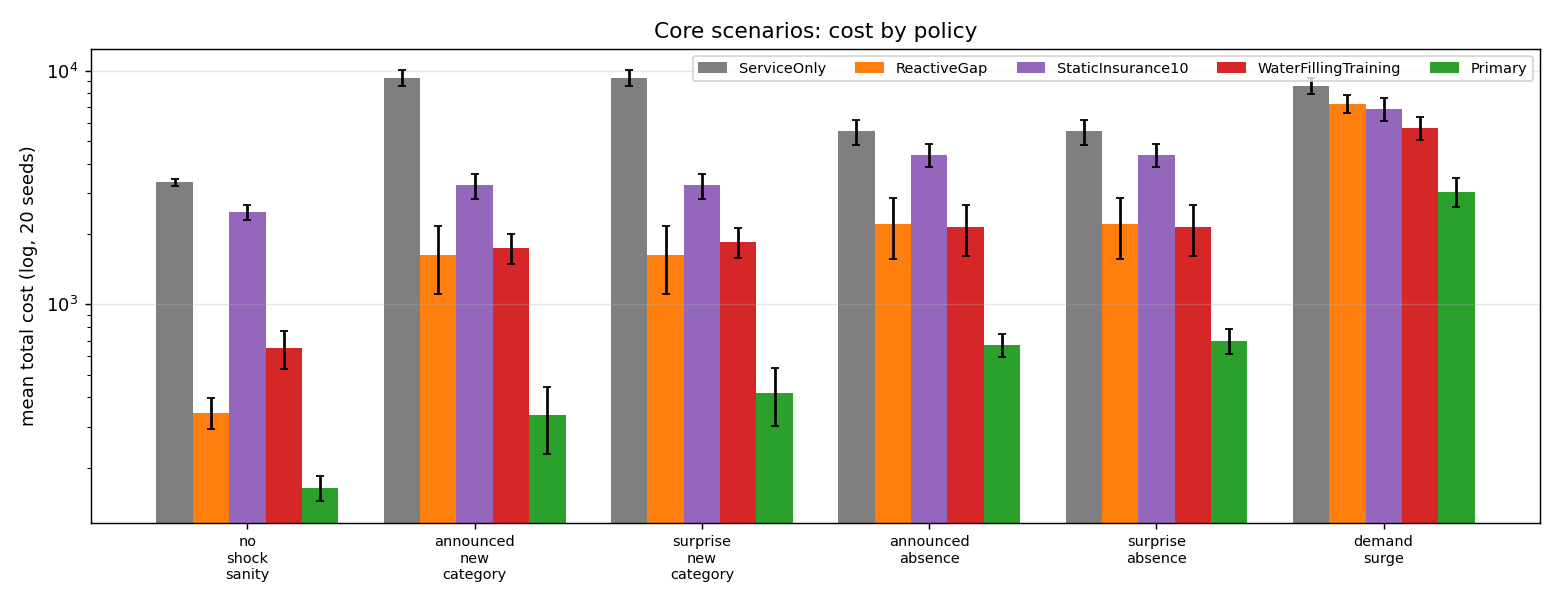}
\caption{Core scenarios: mean cost by policy (log scale, 20 seeds).}
\label{fig:p3core}
\end{figure}

\begin{figure}[t]
\centering
\includegraphics[width=0.49\linewidth]{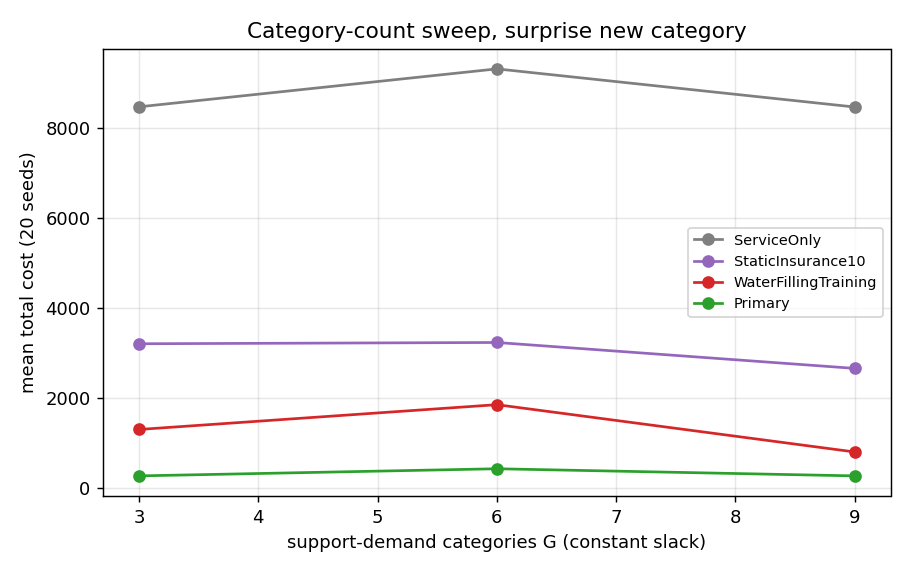}
\hfill
\includegraphics[width=0.49\linewidth]{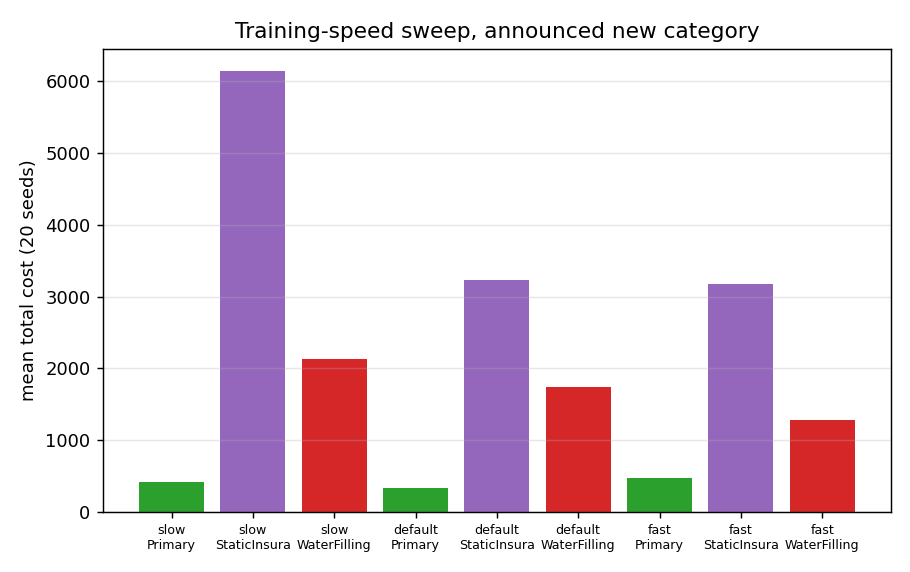}
\caption{Category-count sweep (left) and training-rate sweep (right) on the
surprise/announced new-category shocks.}
\label{fig:p3sweeps}
\end{figure}

Figures~\ref{fig:p3core}--\ref{fig:p3sweeps} summarize the core costs, the
category-count sweep, and the training-rate sweep.

\section{Discussion and Limitations}
\label{sec:p3discussion}

\paragraph{Reading the results carefully.}
The benchmark's value is the regimes it makes measurable. A service-only
policy collapses whenever qualification is the binding resource ---
structurally on new support categories (nobody is qualified; eligibility is
hard) and gradually under decay, where unmaintained qualifications lapse
mid-horizon. The notable finding is a regime map, not a dominance claim. In the
reaction-feasible regime --- where a new qualification can be acquired within
the controller's horizon and recoverable window, including the core instance ---
the closed-loop controller dominates, and we take pains to establish this is not
a weak-baseline artifact: the static-insurance plans are lean, monotone in their
budget, and exactly feasible, they pre-qualify the shock category, and a frozen
adversarial suite (\texttt{adversarial\_validation\_*}) adding a shock-focused
static plan and sweeping disruption windows down to two periods, training rates
across a $4\times$ range, and low slack finds no static win across $158$ paired
cells; a deadline re-scoring (\texttt{deadline\_backlog\_*}) shows that lead does
not depend on backlog being recoverable. The perfect-foresight reference adds
nothing over the primary on announced shocks --- it is a peer, not an upper
bound --- and isolates a $24\%$ value of foresight only under surprise. But we
deliberately sought the other side of the boundary and found it: a frozen
boundary search (\texttt{boundary\_search\_*}) pushing the training lag past the
controller's horizon (cold start, deep retraining) shows lean static insurance
winning by up to $2\times$ on long windows. That win is structural --- the
perfect-foresight peer ($H{=}6$) ties the primary exactly and loses identically,
because no six-period controller can acquire a qualification that takes longer
than six periods to learn. This is exactly the manufacturing-studies regime
(pre-bought capacity wins when reaction is structurally too late), reached here
through a long training lag rather than an unrecoverable capacity transient. Two
further regimes complete the map: a reactive trainer that starts after onset
wastes effort and is worst of all, and under structural capacity insufficiency
($\rho<1$) coverage collapses for every policy and the choice stops mattering.
Forecast privilege is made explicit by construction: every policy declares its
information set, surprise variants mask the forecast until onset, and the oracle
is labeled and never competes.

\paragraph{What the attribution chain establishes.}
The ServiceOnlyMPC $\to$ MaintenanceMPC $\to$ full-controller chain holds
receding-horizon service allocation constant and varies only training
eligibility, separating three mechanisms with direct metric support from
the within-episode counters: qualification \emph{maintenance} against decay,
\emph{requalification} of lapsed cells (invisible to terminal counts by
construction), and \emph{greenfield acquisition} for categories nobody
initially serves. The same counters discipline the narrative: a policy that
wins on cost while acquiring zero new qualifications is winning on
allocation, not capability.

\paragraph{Access dispersion, carefully.}
The Jain indices over per-category coverage and per-staff training hours,
with their minimum floors, expose concentration: policies that achieve
average coverage by systematically under-serving one category, or that
concentrate training on few staff members. These are operational dispersion
metrics over synthetic categories and staff groups. No demographic
attributes exist anywhere in the model, and no fairness or equity claims are
made or implied.

\paragraph{Limitations.}
Everything here is stylized computational evidence. The skill model is a
two-parameter abstraction (linear gain, geometric decay, hard threshold);
demand categories are synthetic and student-free; the instance is small
(eight staff, up to nine categories, one institution); absences and surges
are simple windows; forecast modes are masks over scenario means rather
than learned forecasters. Excluded by design: timetabling, rooms,
transportation, student-level assignment, protected attributes, compliance
logic, learning outcomes, hiring, and procurement. The MILP controller runs
with a deterministic node budget; its solutions are good incumbents, not
proven optima, and solver scalability beyond this instance is untested. The
static-insurance plans are informed comparators (they know which
qualification slots exist, though never whether demand will arrive), and
the oracle reference is a perfect-foresight peer (a receding-horizon $H{=}6$
incumbent within a time budget), not an upper bound. None of the policies
constitutes deployment advice: the framework identifies which institutional
data — real demand categories, training durations, decay rates, absence
patterns — would be needed before any real-world use.

\paragraph{Outlook.}
The natural next steps follow the benchmark discipline: richer qualification
structures (many-to-many category--skill maps where blanket insurance cannot
scale), stochastic or scenario-based controllers that price hidden shocks,
calibration of demand shapes and training durations to public education
statistics \cite{OECD2025EducationAtGlance}, and user studies of the Studio
interface with educational operations leaders — each an extension of the
testbed, not a claim the current paper makes.

\section{Conclusion}
\label{sec:p3conclusion}

We introduced a benchmark specification and decision-support framework for
qualified educational capacity planning: a stylized, single-institution
service system in which support demand is heterogeneous by category,
services are not storable, staff qualification is a dynamic state with hard
threshold eligibility and decay, and training is a control action that
consumes the same staff hours current service needs. The benchmark ships
six seed-controlled scenario families with announced and surprise
information regimes, an exact-feasibility discipline with declared
per-policy information sets, within-episode requalification and greenfield
counters, access-dispersion metrics, replay checksums, and paired
statistics; a policy suite spanning service-only, reactive, static-
insurance, water-filling, and rolling-horizon MILP control with an
attribution chain and a labeled oracle reference; and EduCapacity Studio,
a scenario-analysis interface whose export/replay loop reproduces results
exactly.

The empirical picture is a regime map, governed by whether a newly required
qualification can be acquired within the controller's reaction reach. In the
reaction-feasible regime --- which includes the core instance and a frozen
$2{,}620$-episode adversarial suite (a shock-focused static plan, windows down to
two periods, a $4\times$ training-rate range, low slack) --- the closed-loop
controller dominates over all $158$ paired cells, and we establish this is
legitimate rather than a weak-baseline artifact. But when the training lag is
pushed past the controller's horizon (a cold start with deep retraining), a
$4{,}420$-episode boundary search finds lean static insurance winning by up to
$2\times$ on long windows; the win is structural, since the perfect-foresight
peer ties the controller exactly and loses identically. A reactive trainer that
starts too late is worst of all, and under structural capacity insufficiency no
policy choice matters; backlog perishability shifts the boundary without erasing
either regime. This recovers --- rather than contradicts --- the
manufacturing-studies regime split: pre-positioning wins where reaction is
structurally too late, reached here through a long training lag relative to the
controller's horizon. The benchmark's
contribution is to make these regimes --- and the data one would need to
locate a real institution within them --- measurable, reproducible, and
inspectable, without claiming validated educational impact, compliance, or
student-level recommendations. Extensions in
qualification structure, uncertainty-aware control, public-data
calibration, and interface evaluation are left as future work on top of the
released specification.

\section*{Funding}
This research received no external funding.

\bibliographystyle{plain}
\bibliography{references}

\end{document}